\documentclass[multphys,vecphys]{svmult}

\usepackage{makeidx}
\usepackage{graphicx}
\usepackage{multicol}
\usepackage[bottom]{footmisc}

\makeindex

\begin{document}

\title*{The Cosmogony of Super-Massive Black Holes}

\author{Wolfgang J.\ Duschl\inst{1,2}\and
Peter A.\ Strittmatter\inst{2}}

\institute{Institut f\"ur Theoretische Astrophysik, Zentrum f\"ur
Astronomie der Universit\"at Heidelberg, Albert-Ueberle-Str. 2,
69120 Heidelberg, Germany
\texttt{wjd@ita.uni-heidelberg.de}
\and
Steward Observatory, The University of Arizona, 933 N. Cherry Ave.,
Tucson, AZ 85721, USA
\texttt{pstrittmatter@as.arizona.edu}}

\maketitle

\section{Introduction and Motivation}
\label{WJD:intro}

Two recently found  properties of (super-massive) black holes
(SMBHs) in the centers of active galaxies shed new light on their
formation and growth:

\begin{itemize}

\item The luminosities of the quasars with the largest redshifts
indicate that central BH masses of $> 10^9\,\mathrm{M}_\odot$ were
already present when the Universe was less than $10^9\,$years old
(Barth et al.\ 2003). These masses are lower limits as they are
based on the assumption that the BHs accrete at their Eddington
limit. There is no indication that the (majority of the) sample of
highest-redshift-quasar luminosities is afflicted by amplification
by gravitational lensing (White et al.\ 2005).

\item Surveys in the X-ray regime (Hasinger et al.\ 2005) and in
the optical/UV regime (Wolf et al.\ 2003) show a strong luminosity
dependence of the redshift at which the active galactic nuclei (AGN)
space density peaks: The lower the AGN luminosity, i.e., the smaller
the BH mass, the later in the evolution of the Universe the
co-moving space density of these AGN peaks. In other words, it takes
BHs of a lower {\it final\/} mass longer to reach that mass than BHs
of a larger final mass. This is also supported by a comparison of
the local and the derived accretion mass function of SMBHs (Shankar
et al.\ 2004).

\end{itemize}

\noindent In this contribution we report results of a project
investigating the growth of SMBHs by disk accretion. We find that
both above-mentioned phenomena can be explained in the framework of
such a model.

\section{Black Hole Formation and Growth in
Galactic Centers} \label{WJD:scenario}

Our model involves two basic processes, namely: (a) galaxy mergers
resulting in aggregation of material into a self-gravitating disk
around the galactic center (GC); and (b) subsequent disk accretion
into a central BH. The issue in the past has been whether enough
disk material could be accreted in the available time, a problem
that has become steadily more acute as increasingly luminous quasars
have been found at great redshifts.

In regard to (a), we envisage that the interaction between two (gas
rich) galaxies leads to the rapid formation of a circum-nuclear
gaseous disk. This process is expected to occur on a dynamical time
scale. The mass and spatial extent of this disk will depend on the
mass and gas content of the interacting galaxies and on the impact
parameters of the collision. The process will thus result in a range
of disk masses and extents.

For our subsequent reasoning it is important to note that fairly
early in the evolution of the Universe ($z > 1.5 \cdots 3$) massive
galaxies were already in place (Chen \& Marzke 2004, Glazebroke et
al.\ 2004, van Dokkum \& Ellis 2003). Moreover, there is mounting
evidence (e.g., Dunlop et al.\ 2003, S\'anchez et al.\ 2004) that
interactions between and mergers of galaxies trigger nuclear
activity (e.g., McLeod \& Rieke 1994a\&b, Bahcall et al. 1997,
Canalizo \& Stockton 2001, S\'anchez et al. 2005, Sanders 2004).

Thus for a {\it major merger\/} the resulting gaseous disk mass may
well contain $> 10^{10}\,\mathrm{M}_\odot$ within a few hundred
parsecs of the GC. This basic process has been demonstrated by
numerical simulation (e.g., Barnes 2002, Barnes \& Hernquist 1996 \&
1998, Iono, Yun, \& Mihos 2004) and provides the essential initial
conditions for our analysis. The resulting nuclear disk will, of
course, be subject to viscous dissipation causing an inward flow of
material towards the GC where it is potentially available for
accretion into a BH.  With the initial mass and radius parameters
discussed above, such a disk must inevitably be self-gravitating, at
least initially. In these circumstances, the disk accretion time
scale and, hence, the growth times and limiting mass of the putative
BH, depend on both the mass and extent of the initial disks.

We also note that, initially, the disk may provide mass to the BH at
a rate higher than is allowed by the Eddington limit. Thus in such a
case initially the BH growth rate is defined by the Eddington limit,
with the rest of the material presumably driven from the system (or
at least the proximity of the BH) by radiation pressure. The peak
luminosity will occur roughly when the rate of supply of material
from the disk equals the Eddington limit for the current BH mass
(higher for higher accretion rate).

\section{Evolution of Self-Gravitating Accretion Disks and the
Growth of Black Hole Masses}\label{evol}

We carried out numerical simulations modelling the evolution of
initially self-gravitating accretion disks and the ensuing growth of
the central BH. Our model disks are geometrically thin and
rotationally symmetric, with the following modifications with
respect to standard accretion disk models:

\begin{itemize}

\item We allow for a disk mass which is not necessarily small
compared to the mass of the central BH, i.e., we do not assume a
Keplerian rotation curve in the disk but solve Poisson's equation.

\item We use the generalized viscosity prescription by Duschl et al.
(2000; $\beta$-viscosity).

\item We take the Eddington limit into account. Mass flow above the
Eddington limit is assumed to be lost from the system.

\end{itemize}

\noindent Our numerical code is based on an explicit
finite-difference scheme. For further details of the modelling, we
refer the reader to an upcoming paper (Duschl \& Strittmatter, {\it
in prep.\/})

\subsection{A Reference Model}

As a Reference Model, we define an accretion diks with the following
set of parameters:

\begin{itemize}

\item Inner radius of the disk: $s_{\rm i} = 10^{16.5}\,$cm

\item Outer radius of the disk: $s_{\rm o} =
10^{20.5}\,\textrm{cm} \approx 10^2\,$pc

\item Initial disk mass: $M_{{\rm d},0} =
10^{10}\,\textrm{M}_\odot$

\item Initial surface density distribution: $\Sigma_0 \left( s
\right) \propto s^{-1}$.

\item Seed black hole mass: $M_{{\rm BH},0} =
10^6\,\textrm{M}_\odot \ll M_{{\rm d},0}$

\item Viscosity parameter: $\beta = 10^{-3}$.

\end{itemize}

\subsection{The Evolution of the Reference Model}

\begin{figure}

\centering

\includegraphics[width=\textwidth]{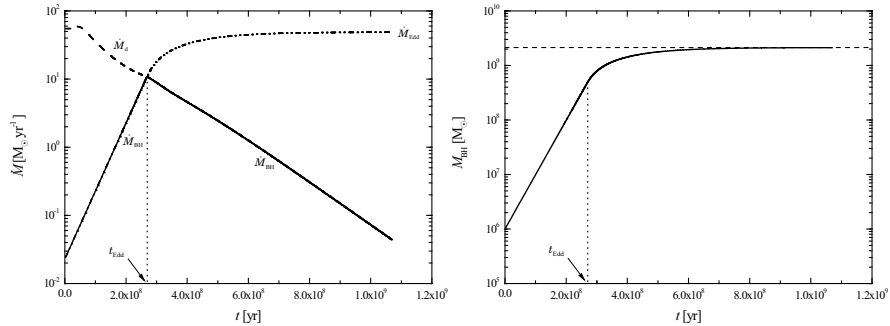}

\caption{The evolution of the mass flow rate (left panel) and the BH
mass (right panel) for the reference model.}

\label{WJD:fig1}

\end{figure}

The evolution of the mass flow rate of the reference model is shown
in the left panel of Fig.\ \ref{WJD:fig1}. The right panel shows the
corresponding evolution of the BH mass. The zero-point of the time
is the point at which, as a consequence of a galaxy-galaxy
interaction, a massive nuclear accretion disk has been established.
One can clearly discern two phases of the accretion process: From
the beginning of the evolution to $t_{\rm Edd} \sim 2.7\cdot
10^8\,$years the evolution is dominated by the Eddington limit: The
disk delivers mass at a larger rate (broken line; $\dot M_{\rm d}$)
than the BH can accrete due to the Eddington limit (dash-dot-dotted
line; $\dot M_{\rm Edd}$). For times $t < t_{\rm Edd}$ the growth
rate of the BH, $\dot M_{\rm BH}$, is subject to the Eddington
limit, i.e., $\dot M_{\rm BH}\left( t < t_{\rm Edd} \right) = \dot
M_{\rm Edd}$. For $t > t_{\rm Edd}$, however, both the mass of the
BH has become so large and the mass flow rate from the disk has
dropped by so much that $\dot M_{\rm d}$ now has fallen below $\dot
M_{\rm Edd}$ and all the mass delivered by the disk can be accreted:
$\dot M_{\rm BH}\left( t > t_{\rm Edd} \right) = \dot M_{\rm d}$.
For the following few $10^8\,$years the free accretion rate,
however, is still large enough to make the BH grow at a fast rate.
This is slowed down considerably only after another $\sim 3.5\cdot
10^8\,$years by when the accretion rate has fallen by approximately
one and a half orders of magnitude. From now on the BH grows only
slowly; it has almost reached its {\it final\/} mass of $2.1\cdot
10^9\,\textrm{M}_\odot$ (broken line in the right panel).

\subsection{Variations of the Initial Physical Setup: The Black Hole
Growth Time }

\begin{figure}

\centering

\includegraphics[width=0.6\textwidth]{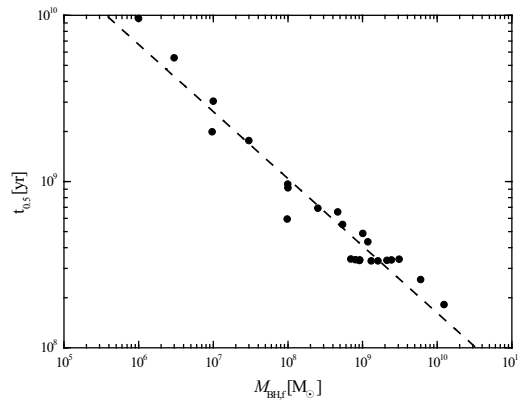}

\caption{The BH growth time scale $t_{0.5}$ as a function of the
final BH mass $M_{\rm BH}$.} \label{WJD:fig2}

\end{figure}

As an example of the influence of the variation of the initial
physical setup, we show in Fig.\ \ref{WJD:fig2} the growth time
scale $t_{0.5}$ of BHs where the initial disk mass and the inner
disk radius have been changed, while all the other parameters of the
reference model remained unaltered. $t_{0.5}$ is defined as the time
at which the BH has reached half its final mass. In all our models,
at this time the accretion rate, and thus the accretion luminosity
have already fallen considerably below their maximum value. It is
noteworthy that for BH masses in the realm of our GC, the growth
times reach values comparable to the Hubble time.

\section{Discussion and Outlook}
Massive accretion disks seem to have the required properties to
explain the observations described at the beginning of this
contribution: The BH mass growth is quick enough to account for the
inferred masses in the highest-redshift quasars, and the evolution
time is an inverse function of the final BH mass\footnote{For a
presentation of the entire set of model calculations and for a more
exhausting discussion of their results, we refer the reader to an
upcoming paper (Duschl \& Strittmatter, {\it in prep.\/}).}. We
expect that the evolution of the Universe as a whole will even
emphasize the latter effect: In the early Universe, galaxy mergers
and collisions were much more frequent than they are in today's
Universe making high initial disk masses more likely at higher
redshifts. For a detailed comparison to the observed luminosity
functions (e.g., Hasinger et al.\ 2005) this cosmological evolution
of the initial conditoions has to be taken
into account.\\

\noindent {\bf\it Acknowledgements.} We benefitted very much from
discussions of the topic with G.\ Hasinger and S.\ Komossa
(Garching) and A.\ Burkert and T.\ Naab (Munich). This work is
partially supported by the German Science Foundation DFG through the
Collaborative Research Center SFB439 {\it Galaxies in the Young
Universe\/}.



\printindex
\end{document}